\documentstyle[floats,prb,aps]{revtex}

\parskip=\medskipamount
\begin{document}
\input{epsf}
\draft

\twocolumn[\hsize\textwidth\columnwidth\hsize\csname @twocolumnfalse\endcsname

\title{Spectral statistics near the quantum percolation threshold
}

\author
{Richard Berkovits and Yshai Avishai$^{\dag}$}

\address{
The Minerva Center for the Physics of Mesoscopics, Fractals and Neural 
Networks,\\ Department of Physics, Bar-Ilan University,
Ramat-Gan 52900, Israel}

\address{
$\dag$also at Department of Physics, Ben-Gurion University, Beer-Sheva, Israel}

\date{\today}
\maketitle

\begin{abstract}
The statistical properties of spectra of a 
three-dimensional quantum bond percolation
system is studied in the vicinity of the metal insulator transition.
 In order to avoid the influence of small clusters,
only regions of the spectra in which the density of states is rather smooth
are analyzed. Using finite size scaling hypothesis, the critical 
quantum probability for bond occupation is found to be $p_q=0.33\pm.01$
while the critical exponent for the divergence of the
localization length is estimated as $\nu=1.35\pm.10$. This later figure
is consistent with the one found within the
 universality class of the standard 
Anderson model.
\end{abstract}

\vskip2pc]
\narrowtext

The present work is concerned with level statistics in an 
Anderson type quantum percolation model. More specifically, 
we consider a single particle  in a three-dimensional
lattice with binary distribution of bonds and analyze (numerically)
the distribution $P(s)$ of adjacent level spacings $s$ 
for bond occupation probabilities close to the 
critical one (which marks the metal insulator transition).

Level statistics in quantum systems and its relation to
random matrix theories constitutes an important tool for
understanding the underlying physics \cite {Ref1a}. In particular,
correlations between energy eigenvalues of a single 
quantum particle interacting with random impurities
in the diffusive regime are consistent with the predictions of 
Gaussian matrix ensembles \cite{Ref2a,Ref2b,Ref2c,Ref2d,Ref2e}. 
Recently, it became clear
that in the vicinity of a metal insulator transition 
(provided it exists in such systems) there is a distinct kind of
level statistics\cite {Ref3a}. In this novel statistics, the critical exponent
for the divergence of the localization length appears in numerous
expressions for the various correlations\cite {Ref3a,Ref3b,Ref3c,Ref3d}. 
Hence, it is difficult
to perceive a random matrix theory which adequately describe this
critical statistics, although some progress has been recorded in
this direction \cite {Ref3e}. 
One of the clearest indications for the existence of a different
statistics in the neighborhood of the metal insulator 
transition is displayed in the behavior of
the nearest level spacing distribution $P(s)$, which, for large level
spacing $s$, falls off slower than Gaussian \cite {Ref3c}. 
This is found to be the case both for the Anderson metal insulator
transition in three dimensions \cite {Ref5a,Ref5b} as well as for
the Hall transition in two dimensions \cite {Ref6a}.

One of the motivations for studying level statistics in a quantum
percolation model is related to the question of whether it
belongs to the same universality class of the Anderson model
with site disorder \cite {Ref7,Ref7'}. The answer to this question is by no
means clear, despite the fact that quantum percolation can
be regarded as a special variant of the general Anderson model
\cite {Ref7a}. For example, in some quantum percolation
models, the value of the critical exponent $\nu$ for the divergence
of the localization length, as can be deduced from the transmission of 
the system, is found to be smaller than that
of the Anderson model \cite {Ref8a,Ref8b}. Our analysis 
suggests that for a tight binding model
the critical exponent (as can be deduced from the level statistics)
for site disorder and that
for quantum (bond) percolation are nearly identical.

Another motivation
(upon which we will not elaborate in this work) concerns with the
fractal nature of the wave function near the critical point. In
particular, if the critical quantum probability for bond occupation
(denoted hereafter as $p_{q}$) is only slightly higher than the
classical one (denoted hereafter as $p_{c}$) then the critical
wave functions live on a fractal object, and the geometrical
fractal dimension becomes relevant.

Let us start by introducing the quantum percolation model
and then explain how the nearest level spacing distribution
is computed. Our calculations are based on a tight binding
Hamiltonian,
\begin{equation}
H= \sum_{\langle ij \rangle} (t_{ij} a_i^{\dag} a_j + h.c),
\label{Hamiltonian}
\end{equation}
where $\langle ij \rangle$ denotes nearest neighbors.
The hopping matrix elements $t_{ij}$ are independent random
variables which assume the values $1$ or $0$ with probabilities
$p$ and $q=1-p$ respectively. The underlying lattice is a three
dimensional cube of length $L$ with periodic boundary conditions. 
The missing bond probability $q$ plays the role of disorder
strength. For each realization $k$ of bond occupation probability, 
$p$, the above 
Hamiltonian is diagonalized exactly, yielding a sequence of
eigenvalues $E_{n}^{k}$, $n=1,2,...L^{3}$. This sequence is calculated
for $N$ different realizations, where $N =3000,1400,750,450,300$
for the corresponding different sample sizes $L=7,9,11,13,15$.
This corresponds to $10^6$ eigenvalues for each sample size.

\begin{figure}
\centerline{\epsfxsize = 3in \epsffile{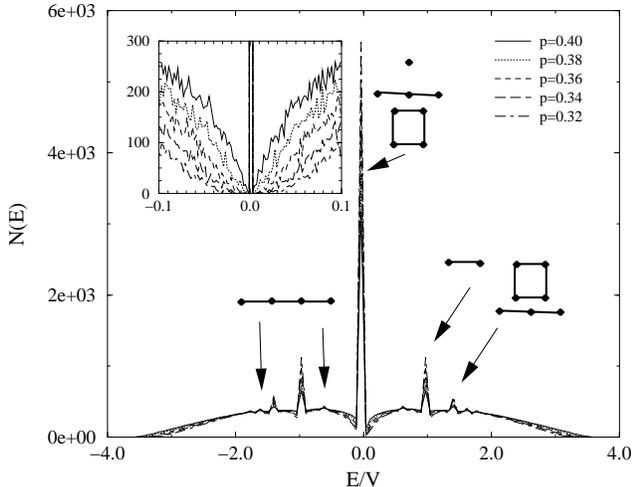}}
\caption{The DOS for $L=13$ as function of energy for different bond
occupation values. The connection between various small clusters and
peaks in the DOS are indicated in the figure. In the inset, an enlargement
of the region around $E=0$ is presented.
\label{fig1}}
\end{figure}

The average density of states (DOS)
for $L=13$ as a function of $p$ is presented in
Fig. \ref{fig1}. The most noticeable feature is the appearance
of a series of sharp peaks in the average DOS which increase as $p$
decreases. This feature was already noted in Ref. \onlinecite{Ref9a}
where the DOS for a quantum percolating system was calculated
using the Sturm sequence method.
The origin of these peaks is the formation of small disconnected
clusters of sites in the sample. For example, a single site with no connecting
bonds to neighboring sites always contributes an eigenvalue 
$\varepsilon=0$. The probability for such a site is equal to $(1-p)^6$,
therefore one expects a contribution of $L^3(1-p)^6$ eigenvalues equal to
zero to the spectrum. This is in agreement\cite{rem1} with the observed height
of the central peak in Fig. \ref{fig1} (the bin size is 0.072) and with its
variation as function of $p$. Another prominent feature is the
appearance of a gap in the DOS which depends on $p$ around the central peak
\cite{Ref9a} which may be seen in the inset of Fig. \ref{fig1}.
A cluster of two sites connected by a
bound has a probability of $p(1-p)^{10}$ to appear and contributes
eigenvalues $\varepsilon=\pm1$ to the spectra.
Similarly, clusters of three sites contribute $\varepsilon=0,\pm\sqrt{2}$
and clusters of four sites contribute $\varepsilon=0,0,\pm\sqrt{2}$ if
all the sites are connected among themselves and 
$\varepsilon=\pm(3\pm\sqrt{5})/2$ if only three bonds are present.
It is interesting to note that gaps seem to develop also around these
peaks.

Here we face the question of how to study a spectrum
for which some of the levels form degenerate clusters. Indeed,
one can apply the various statistical measures of level
statistics only if the density
of states is smooth. Looking at Fig. \ref{fig1}, one may
concentrate on three such regions centered 
around (I) $E=\pm 0.4$, (II) $E=\pm0.8$,
(III) $E=\pm 1.2$ (the spectrum for an odd $L$ with periodic 
boundary conditions is not symmetric).
In each region a fixed number of levels are taken ($15,31,
57,95,145$ for $L=7,9,11,13,15$) and the spectrums unfolded by the
usual procedure, i.e., $x_{i+1} = x_{i} + s_{i}$ and
$s_{i} = n (E_{i+1}-E_{i})/(E_{i+\lfloor n/2 \rfloor+1}-
E_{i-\lfloor n/2 \rfloor })$. In the data presented
here $n=13$ is used, but no significant difference is seen for $n=9$.
Within these guidelines,
the distribution of adjacent level spacings for each region, sample size $L$ 
and bond probability $p$ is then calculated.

\begin{figure}
\centerline{\epsfxsize = 2in \epsffile{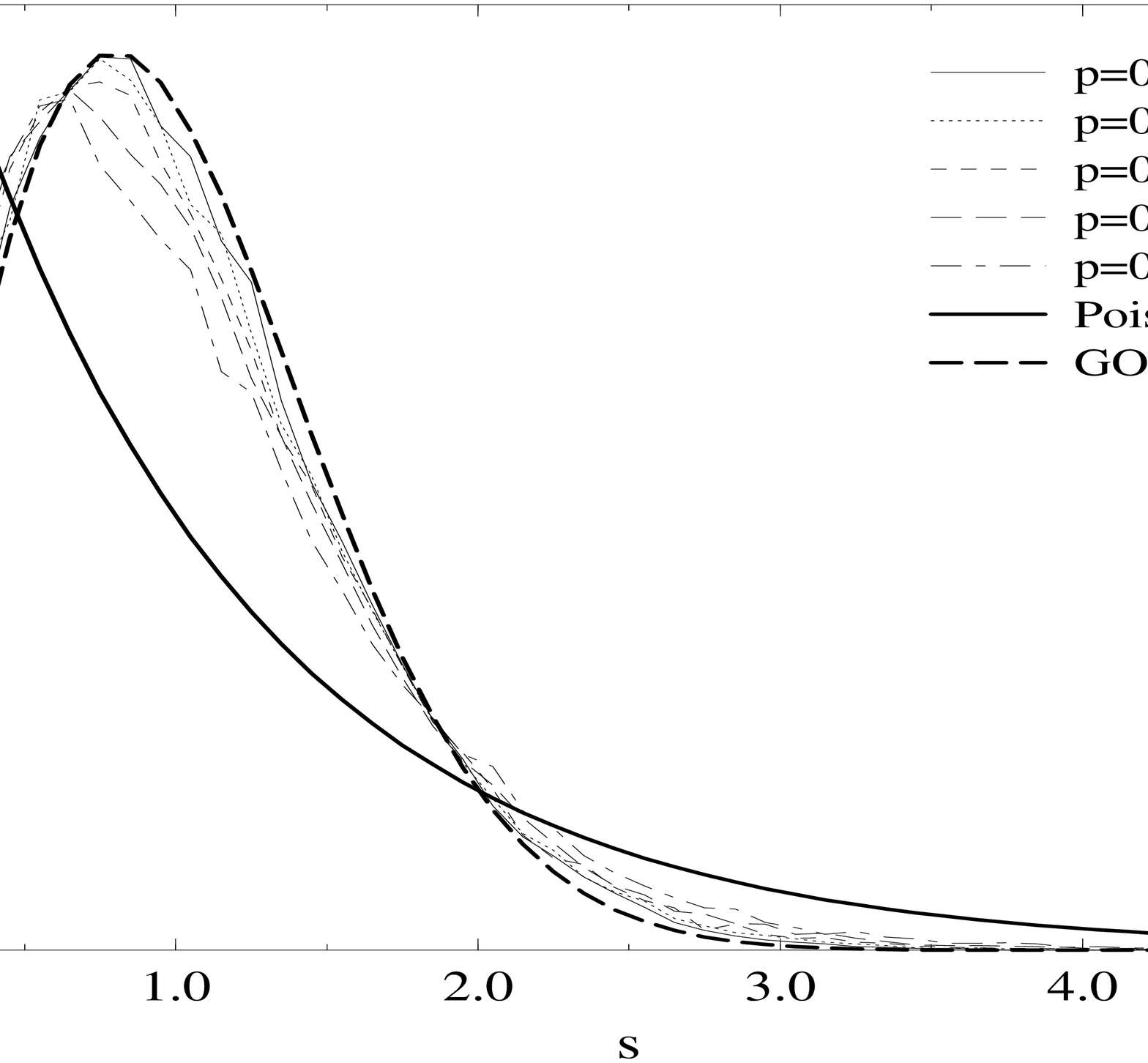}}
\caption{The level spacing distribution for $L=13$. One can see the transition
from a GOE distribution (indicated by the thick full line) towards
a Poisson distribution (indicated by the thick dashed line) as $p$ decreases.
\label{fig2}}
\end{figure}

\begin{figure}
\centerline{\epsfxsize = 2in \epsffile{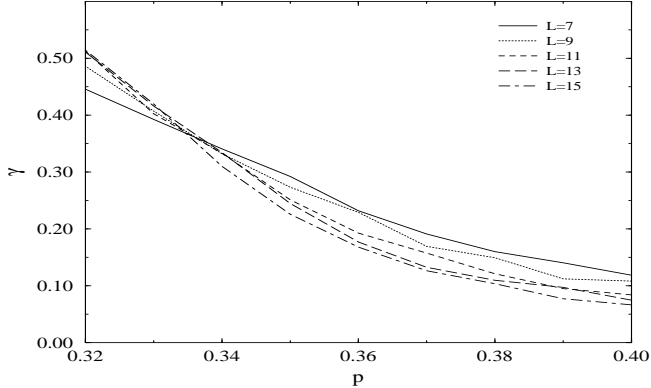}}
\caption{The scaling function $\protect \gamma$ as function of $p$ for 
different sample sizes for levels around $\protect E=\pm0.4$. A clear
convergence of all curves at $p_q\sim0.33$ can be seen, as well as 
the expected change in the size dependence of $\protect \gamma$.
\label{fig3}}
\end{figure}

A plot of P(s) as function of the bond occupation 
probability for $L=13$ is displayed
in Fig. \ref{fig2}. It can be clearly seen that the expected transition from
a Wigner like behavior for large $p$ to a Poisson behavior for small $p$
is manifested. One should also note that all curves seem to intersect at
$s\sim2$, which reminds us of the situation for the Anderson transition with
on-site disorder\cite{Ref3a}.
As has been shown in Ref. \onlinecite {Ref3a} a very convenient way to
obtain the mobility edge as well as the critical exponent of the transition
$\nu$ is to study the parameter $\gamma(p,L)$ defined as
\begin{equation}
\gamma(p,L)= {{\int_2^\infty P(s) ds - e^{-\pi}}\over{e^{-2}- e^{-\pi}}},
\label{Gamma}
\end{equation}
which characterizes the transition from Wigner to Poisson. 
Denoting by $\xi(p)$ the localization length, this function
is expected to show a scaling behavior $\gamma(p,L) = f(L/\xi(p))$ which
in the vicinity of the critical quantum bond probability $p_q$ is expected to
behave as\cite{Ref3a} 
\begin{equation}
\gamma(p,L)= \gamma(p_q,L) + C \left| {{p}\over{p_q}} - 1 \right| L^{1/\nu},
\label{Tran}
\end{equation}
where $C$ is a constant. In Fig. \ref{fig3} curves of $\gamma(p,L)$ for 
different sample sizes $L$ are plotted for levels 
in the first energy domain. 
It is noticed that the curves cross at a single point at which
the order of heights with respect to $L$ is reversed.
This is an indication for the existence of
finite size one parameter scaling behavior.
Similar situation prevails also in regions II and III.

Based on finite size one parameter scaling analysis,
the procedure for calculating
the critical bond probability, as well as the critical exponent
goes as follows. The quantity $\gamma(p,L)$ is calculated
for many pairs $(p_{i},L_{i})$. It is then 
considered as a
certain scaling function $f(x)$ of the scaling variable
$x=L^{1/\nu} (p-p_{q})$. For $x \rightarrow \infty$ 
the system is well inside the diffusive regime and hence
$f(x) \rightarrow 0$. On the other hand For $x \rightarrow -\infty$ 
the system is well inside the insulating regime and hence
$f(x) \rightarrow 1$. Practically, it is useful to shift the
variable $x$ to $y(x)=\frac {x-a-b} {b-a}$ where $a$ and $b$ are
respectively the minimum and maximum values assumed by $x$.
Evidently, $y(x)$ ranges between $-1$ and $1$. 
 Then one expands $f(x)$ in a series
of Tschebicheff polynomials $T_{n}[y(x)]$
($n=0,1,2,...K$). 
Minimization of the set of differences 
$|f(x_{i})-\gamma(p_{i},L_{i})|$ results in the unknowns
$p_{q}$, $\nu$ and the expansion coefficients (namely, the
scaling function itself). In all cases, it is sufficient
to cut off the number of polynomials at $K=12$.

\begin{figure}
\centerline{\epsfxsize = 2in \epsffile{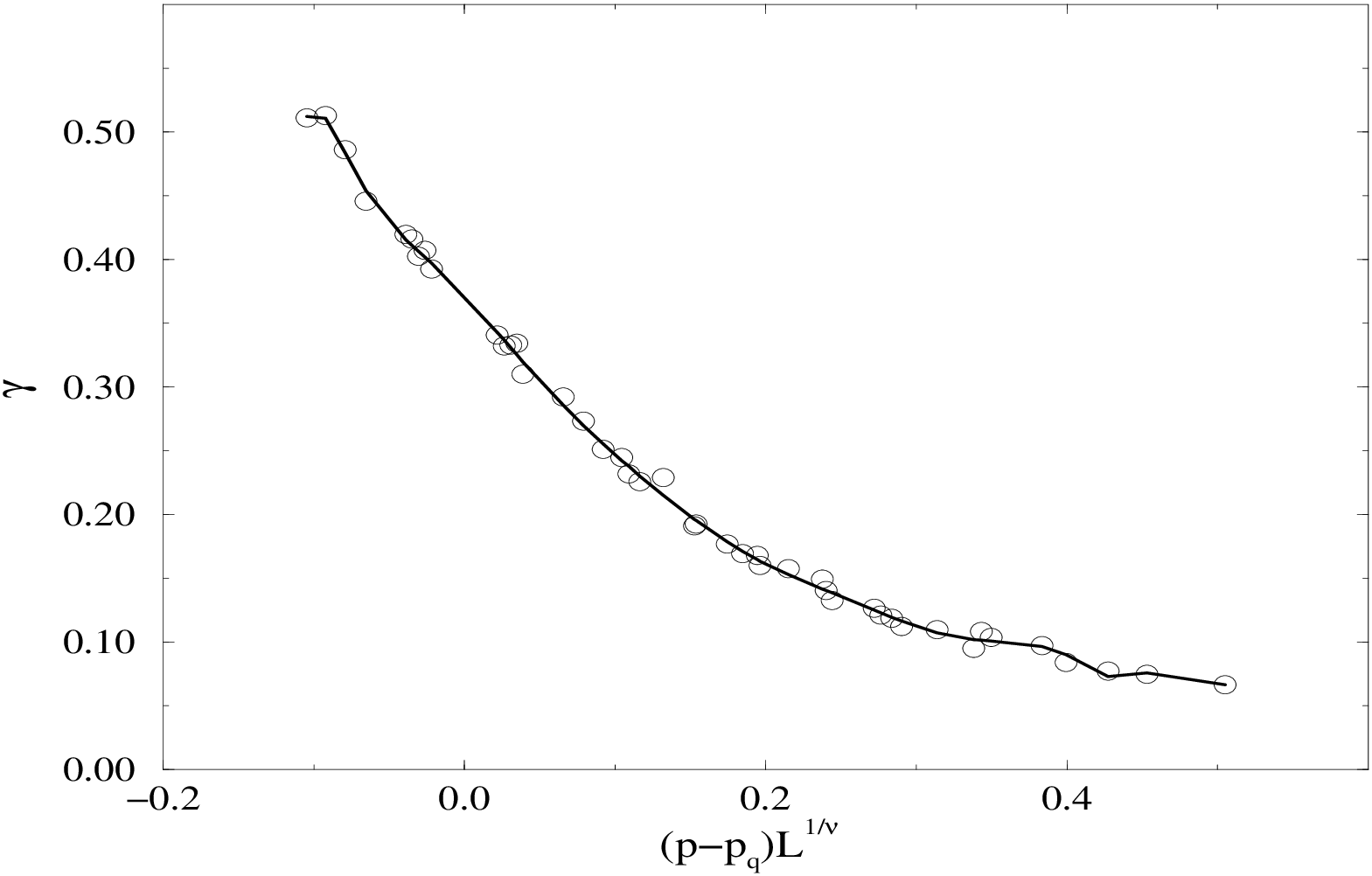}}
\caption{A fit of the numerical data around $\protect E=\pm0.4$ 
to the scaling function represented by the curve.
\label{fig4}}
\end{figure}

The following results are obtained: for region I $p_q=0.335\pm.005$ 
and $\nu=1.32\pm.08$,
for region II $p_q=0.33\pm.005$ and $\nu=1.35\pm.10$ and for region III 
$p_q=0.325\pm.005$ and $\nu=1.35\pm.12$.
As a measure of the quality of the fit the numerical data and
the fitted scaling function are plotted in Fig. \ref{fig4}.
It can be seen that as one might expect $\nu$ is the same for all the 
three regions, while there is a small shift in $p_q$ as $E$ increases.
The value of $p_q$ and its dependence on $E$ 
is in perfect agreement with previous numerical studies of quantum
percolation systems \cite {Ref8b,Ref9a}. On the other hand,
$\nu$ is not consistent with the different values of the critical
exponent obtained for those systems, i.e.,$\nu=0.38$ in Ref. \onlinecite{Ref8a}
and $\nu=0.75$ in Ref. \onlinecite{Ref8b}, but is remarkably
close to its value for the on-site disorder Anderson model
\cite{Ref10a,Ref3a} $\nu=1.5\pm.1$.

\begin{figure}
\centerline{\epsfxsize = 2in \epsffile{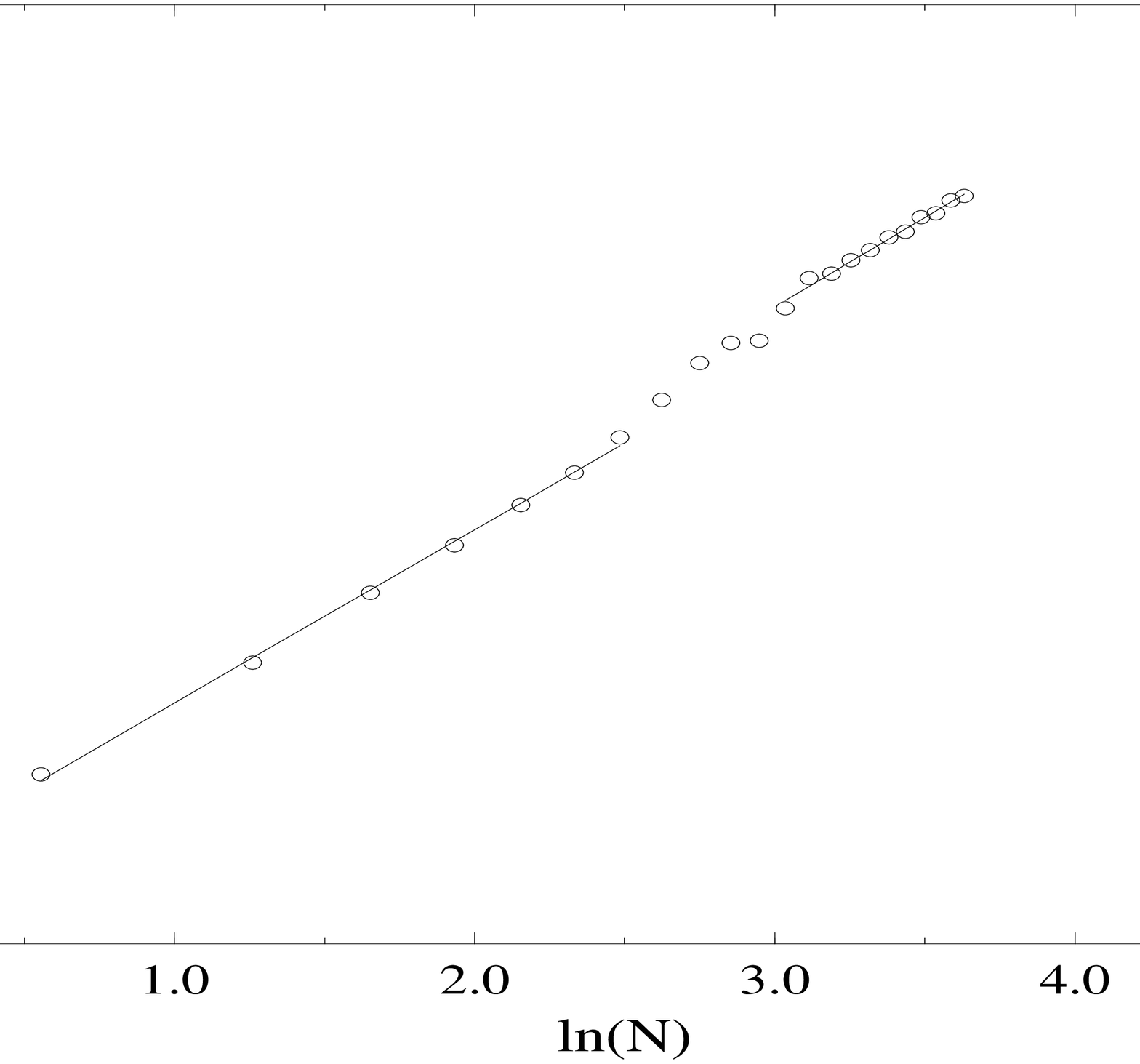}}
\caption{The logarithm of the number variance as function of the 
logarithm of the average number of states for the region around
$\protect E=\pm0.8$. Linear fits were
performed for (i) $\protect 1.5<
\bar N<12$ and (ii) $\protect 20<\bar N<45$. For $\protect 12<\bar N<20$
a non monotonous behavior of the number variance is seen. This behavior
is probably connected to a small peak in the DOS at $\protect E\sim0.83$.
\label{fig5}}
\end{figure}

Another quantity which is sensitive to the critical exponent $\nu$ is the
behavior of the tail of $P(s)$ at
the transition point. According to Kravtsov {\it et. al} \cite {Ref3b}
$\ln(-\ln(P(s)))=(2-\gamma)\ln(s) + {\rm const.}$, 
where $\gamma = 1 - {{1}\over{\nu d}}$. This is not an accurate algorithm
to calculate $\nu$ since it depends on the behavior of $P(s)$ at the
tail of the distribution, for which the statistics is rather poor.
It is important to note that in Ref. \onlinecite{Ref3a} Shklovskii
{\it et. al} predict $\gamma=1$ even in the critical
region with no dependence on $\mu$
which is supported by some recent numerical work on the on-site
Anderson model\cite{Ref11}.
Nevertheless for the quantum percolation model we obtain $\gamma=0.68\pm.16$,
which corresponds to $\nu=1.04^{+1.04}_{-.34}$. A better measure for
$\gamma$ is the number variance $\Sigma^2(\bar N)$ which should behave as
$\Sigma^2(\bar N) \propto \bar N^\gamma$, at least for moderate values
of $N$ in which an additional linear term recently predicted\cite{Ref12}
is not significant\cite{Ref6a}. 
The logarithm of the number variance $\ln(\Sigma^2(\bar N))$
versus $\ln(\bar N)$ is plotted in Fig. \ref{fig5}. Two different
regions for which a linear behavior is observed can be seen (i) $1.5<
\bar N<12$ and (ii) $20<\bar N<45$. In between, a jump in 
$\Sigma^2(\bar N)$ can be seen which might be associated with some
small cluster peak. A linear fit in (i) gives $\gamma=0.74\pm.02$
corresponding to $\nu=1.28^{+.11}_{-.09}$ and in (ii)
$\gamma=0.76\pm.03$ resulting in $\nu=1.39^{+.20}_{-.16}$.
All the above estimations of $\nu$ fall within the range
obtained from the finite size one parameter scaling.

Thus, based on the analysis of various statistical properties of
the quantum percolation spectra, the critical exponent in the well behaved
regions of the spectra is $\nu=1.35\pm.10$. 
This, at least for properties connected to the
energy levels, seems to put the quantum percolation system in the
same universality class as the usual on-site disorder Anderson model.
The previous studies calculated $\nu$ via the transmission of the system
at energies very close to $E=0$. As can be seen in the inset of Fig. \ref{fig1}
the DOS has a very strong $p$ dependence in that region. Thus a large
part of the dependence of the transmission on $p$ is probably due to the change
in the DOS and not because of some changes in the localization properties.
Therefore, it will be very interesting to examine the transmission
in regions of $E$ where the DOS has only a weak dependence on $p$.

We are grateful to D.E. Khmelnitskii and  B. Shapiro
for useful discussions. R.B. would like to thank the 
US-Israel Binational Science Foundation
for financial support. Y.A. thanks the Israeli Academy of Science and
Humanities for financial support.


\begin{references}


\bibitem{Ref1a} M. L. Mehta, {\it Random matrices}
(Academic Press, San-Diego, 1991), and references therein.

\bibitem{Ref2a}L. P. Gorkov and G. M. Eliashberg Zh. Eksp. Teor. Fiz.
{\bf 48}, 1407 (1965) [Sov. Phys. JETP {\bf 21}, 940 (1965)].

\bibitem{Ref2b} B. L. Altshuler and B. I. Shklovskii, Zh. Eksp. \& Teor. Fiz.
{\bf 91}, 220 (1986) [Sov. Phys. JETP {\bf 64},127 (1986)].

\bibitem{Ref2c} U. Sivan and Y. Imry, Phys. Rev. B {\bf 35},
6074 (1987).

\bibitem{Ref2d} S. N. Evangelou and E. N. Economou, Phys. Rev. Lett.
{\bf 68}, 361 (1992).

\bibitem{Ref2e} F. M. Izrailev, Phys. Rep. {\bf 129}, 299 (1990).

\bibitem{Ref3a}B. I. Shklovskii, B. Shapiro, B. R. Sears, P. Lambrianides,
and H. B. Shore, Phys. Rev. B. {\bf47}, 11487 (1993).

\bibitem{Ref3b} V. E. Kravtsov, I. V. Lerner, B. L. Altshuler,
and A. G. Aronov, Phys. Rev. Lett. {\bf 72}, 888 (1994).

\bibitem{Ref3c} A. G. Aronov, V. E. Kravtsov and I. V. Lerner,
JETP LEtt. {\bf 59}, 39 (1994).

\bibitem{Ref3d} A. G. Aronov, V. E. Kravtsov and I. V. Lerner,
Phys. Rev. Lett. {\bf 74}, 1174 (1995).

\bibitem{Ref3e} M. Moshe, H. Neuberger and B. Shapiro,
Phys. Rev. Lett. {\bf 73}, 1497 (1994).

\bibitem{Ref5a} S. N. Evangelou, Phys. Rev. B {\bf49}, 16805 (1994).

\bibitem{Ref5b} E. Hofstetter and M. Schreiber, Phys. Rev. Lett.
{\bf 73}, 3137 (1994).

\bibitem{Ref6a} M. Feingold, Y. Avishai and R. Berkovits,
Phys. Rev. B {\bf 52}, 8400 (1995).

\bibitem{Ref7} Y. Gefen, D. J. Thouless and Y. Imry,
Phys. Rev. B {\bf 28}, 6677 (1983). 

\bibitem{Ref7'} B. Shapiro, Phys. Rev. Lett. {\bf 48}, 823 (1982).

\bibitem{Ref7a} for a recent review see A. Mookerjee, I. Dasgupta and T. Saha,
Int. J. Mod. Phys. {\bf 9}, 2989 (1995), and references therein.

\bibitem{Ref8a} I. Chang, Z. Lev, A. B. Harris, J. Adler, A. Aharony,
Phys. Rev. Lett. {\bf 74}, 2094 (1995).

\bibitem{Ref8b} Y. Avishai and J. M. Luck, Phys. Rev. B {\bf 45}, 1074 (1992).

\bibitem{Ref9a} C. M. Soukoulis, Q. Li and G. S. Grest,
Phys. Rev. B {\bf 45}, 7724 (1992).

\bibitem{rem1} Of course, one must take into account the fact that the
central peak has contributions also from other clusters.

\bibitem{Ref10a} B. Kramer and A. MacKinnon, Rep. Prog. Phys. {\bf56}, 1469
(1993).

\bibitem{Ref11} I. Kh. Zherekeshev and B. Kramer J. Jpn. Applied Phys.
\bf{34}, 4361 (1995).

\bibitem{Ref12} V. E. Kravtsov (preprint, cond-mat/9603166)

\end{references}
\end{document}